\documentclass[prb,twocolumn,amsmath,amssymb,superscriptaddress]{revtex4}
\usepackage[T1]{fontenc}
\usepackage[latin1]{inputenc}
\usepackage{graphics}
\usepackage{amssymb}
\usepackage{amsmath}
\usepackage{overpic}

\newcommand{\be}{\begin{equation}}
\newcommand{\ee}{\end{equation}}
\newcommand{\bea}{\begin{eqnarray}}
\newcommand{\eea}{\end{eqnarray}}

\def\a{\alpha}
\def\b{\beta}
\def\e{\varepsilon}

\def\s{\sigma}

\def\G{\Gamma}
\def\D{\Delta}

\def\ra{\rightarrow}
\def\up{\uparrow}

\def\down{\downarrow}

\def\pd{\partial}

\def\bk{{\bf k}}

\def\bq{{\bf q}}

\def\bA{{\bf A}}

\def\nn{\nonumber}
\def\lb{\label}
\def\pref#1{(\ref{#1})}

\newcount\bozza \bozza=0
\ifnum\bozza=1
\newdimen\shift \shift=-2truecm
\def\lb#1{%
{\label{#1}\rlap{\kern\shift{$\scriptstyle#1$}}}}
\else\def\lb#1{\label{#1}} \fi

\begin{document}

\title{Signature of the Leggett mode in the $A_{1g}$ Raman response:\\ from MgB$_2$ to iron-based superconductors}
 
\author{T. Cea}
\affiliation{ISC-CNR and Dep. of Physics, ``Sapienza'' University of
  Rome, P.le A. Moro 5, 00185, Rome, Italy}

\author{L. Benfatto}
\affiliation{ISC-CNR and Dep. of Physics, ``Sapienza'' University of
  Rome, P.le A. Moro 5, 00185, Rome, Italy}

\date{\today}

\begin{abstract}
The Raman response in a superconductor is a powerful probe to investigate the symmetry of the superconducting gap. Here we show that in a multiband superconductor it also offers the unique opportunity to establish if the driving pairing interaction has an intraband or interband character. In the model with one hole and one electron band the full  gauge-invariant Raman response, obtained by accounting for the fluctuations of both the density and superconducting phase degrees of freedom, is always dominated by the Leggett mode, regardless its nature.  However, while in the case of intra-band dominated pairing the Josephson-like phase fluctuations of the two condensates identify a well-defined peak, as observed in MgB$_2$, for dominant interband pairing the Leggett resonance is pushed at twice the largest gap, resembling apparently a pair-breaking peak. The latter case is in very good agreement with experimental data in iron-based superconductors, suggesting that an interband pairing mechanism should be at play in these systems. These results have also interesting implications for the non-linear optical response probed by means of intense THz fields.
\end{abstract}


\maketitle

\section{Introduction}

The inelastic scattering of light represents one of the most powerful spectroscopic probes for the investigation of the superconducting (SC) state of a system. In a Raman experiment the net effect of the photon in/photon out process can be modelled as a perturbation that couples to the electron density at long-wavelength\cite{klein_prb84,hackl_review}. More specifically, the particle-hole excitations are modulated in the momentum space by a form factor $\gamma_\bk$ dictated by the polarization of the incoming/outgoing light and by the symmetry of the band structure, with $\gamma_\bk$ roughly scaling as the inverse effective mass $1/m^*$ of the carriers. 
In the SC state BCS theory predicts that the Raman charge response displays a threshold and a square-root
singularity at twice the gap $\Delta$ edge, simply reflecting the two-particle density of states\cite{klein_prb84,hackl_review}. This result, along with the polarization dependence of the $\gamma_\bk$ prefactor, allows one to probe the SC gap at different momenta,\cite{hackl_prl94} giving crucial informations on the symmetry of the underlying SC state. For this reason, Raman has been proven to be crucial for the spectroscopic investigation of both conventional and  unconventional superconductors.\cite{hackl_review,hackl_prl94,deveraux_prb95,klein_view}

While standard BCS theory only predicts the existence of a quasiparticle pair-breaking peak, several other collective modes can appear in principle in the Raman response. Physically, they arise because the two particles created when a Cooper pair breaks apart continue to interact before than recombining together. In the usual diagrammatic language for the computation of the Raman response function this can be accounted for by including RPA-like and vertex-corrections-like diagrams due to all the possible intermediate processes coupled to the Raman density fluctuations.\cite{klein_prb84,hackl_review,klein_view} Since also this coupling is dictated by symmetry arguments, Raman scattering becomes also a selective probe of collective fluctuations.
A typical example is provided e.g. by the so-called screening of the Raman response in the  symmetric $A_{1g}$ channel, where $\gamma_\bk$ is more or less a constant. In this case the Raman scattering probes the fluctuations of the electron density,  and the Raman response function coincides with the charge susceptibility at long 
wavelength and finite frequency, that is expected to vanish since it controls the response to an uniform potential. This requirement is violated by the BCS Raman response, but it can be restored by adding\cite{klein_prb84,hackl_review}  the contribution of the charge fluctuations, mediated by Coulomb interactions. Even though this effect is often referred as "Coulomb screening"  the presence of long-range interactions is not necessary to obtain it. Indeed, the vanishing of the dynamic charge response is in general a requirement of charge conservation and gauge invariance.\cite{deveraux_prb95} As it well known,\cite{schrieffer} the BCS approximation violates gauge invariance since it lacks the contribution of all the SC collective modes, that include the fluctuations not only of the charge, but also of the SC phase, that is its conjugate variable\cite{depalo_prb99,randeria_prb00,benfatto_prb04}.

For a multiband superconductor with bands having opposite hole/electron character the phenomenology can be even richer. Considering for example a two-band case, the form factor  $\gamma_\bk^i$ will turn out to be positive/negative on the electron/hole bands, so that  labelling the respective densities 
as $\rho_1$ and $\rho_2$ the Raman response in the  $A_{1g}$ channel can access the relative density fluctuations $\rho_1-\rho_2$  instead of the total ones $\rho_1+\rho_2$. The consequences of this effect have been discussed so far both in the context of the MgB$_2$ superconductor\cite{blumberg_prl07,klein_prb10} and iron-based superconductors (FeSC)\cite{klein_view,boyd_prb09,mazin_prb10,blumberg_prb10}. In the former case it has been argued that the $A_{1g}$ Raman response, proportional to the relative $\rho_1-\rho_2$ density fluctuations between the (electron-like) $\pi$ bands and (hole-like) $\sigma$ bands,\cite{review_mgb2} couples to the relative fluctuations $\theta_1-\theta_2$ of the SC phases in the two bands.  As originally discussed by Leggett\cite{leggett}, this phase mode describes the Josephson-like oscillations  between the phases of the two SC condensates. By using parameter values appropriate for MgB$_2$ the Leggett mode is expected\cite{klein_prb10,sharapov_epjb02} to lie at energies between the two gaps $\Delta_1<\Delta_2$, in agreement with experimental measurements in the $A_{1g}$ channel\cite{blumberg_prl07}.  

Also in the case of FeSC the band structure is made by hole-like pockets and electron-like ones, located at the $\G$ and $X/Y$ points of the 1Fe Brillouin zone, respectively.\cite{si_review16,hirschfeld_review16} However,  in the context of FeSC the role of the Leggett-like mode for the Raman response has been neglected so far, and the main focus has been put instead on the effect of the sign-change of the $\gamma_\bk^i$ factors for the screening in the $A_{1g}$ channel. Indeed, by coupling the Raman response {\em only} to the density fluctuations the authors of Refs.\ \cite{boyd_prb09,mazin_prb10,blumberg_prb10} concluded that the usual screening of the symmetric channel is not operative in FeSC. This result has been used to understand the experiments in FeSC\cite{hackl_prb09,hackl_prl13,hackl_prx14,blumberg_prb16}, that reported a pair-breaking-like peak in the  $A_{1g}$ channel, with an overall intensity as large as in the other non-symmetric channels, confirming apparently the lack of screening in FeSC. 

Even though specific features of the band structure of different materials can be quantitatively relevant, the previous theoretical and experimental results seem to be in contradiction with each other. Indeed, starting from the same general model of a two-band superconductor with one hole and one electron band, in one case (MgB$_2$) the $A_{1g}$ response is claimed to measure only the Leggett mode, in the other case (FeSC) it is claimed to measure only the unscreened quasiparticle response, leading to the usual pair-breaking peak. 

In the present paper we show that this apparent contradiction arises when charge and phase fluctuations are not treated on the same footing. While this does not appear to explicitly violate the total charge conservation, in the multiband case it does not describe properly the relative charge fluctuations between the two bands. By computing the Raman response within a full gauge-invariant effective-action formalism we show that  the $A_{1g}$ Raman response of a  multiband superconductor with bands having opposite character is  always dominated by relative phase fluctuations $\theta_L=\theta_1-\theta_2$ of the SC phases of the two order parameters. However, the resulting Raman susceptibility is drastically different depending on the nature of the pairing interaction. Indeed, while in the case of intra-band dominated pairing, as appropriate for MgB$_2$,\cite{review_mgb2} $\theta_L$ identifies a true collective mode, in the sense that it lies below the largest of the two gaps, when the pairing has dominant interband character  $\theta_L$ identifies a resonance that occurs around twice the largest of the two gaps, with a typical profiles that can be accidentally similar to a standard pair-breaking peak. Besides solving the paradox of existing theoretical results, the comparison between our findings and available experimental data in FeSC provides us with an indirect evidence of a dominant {\em interband} pairing glue in these systems, whose most plausible candidates\cite{si_review16,hirschfeld_review16} are spin fluctuations, which naturally connect hole and electron pockets in FeSC. This result is particularly interesting for those families of FeSC, like e.g. LiFeAs and FeSe, where the role of spin fluctuations on the pairing mechanism is still under debate.\cite{si_review16,hirschfeld_review16,buchner_natmat14,meingast_prl15,borisenko_prl10,vanderbrink_prl11} On a more general perspective, our findings show that  Raman scattering  in multiband superconductors as a fundamental probe not only of the SC gap modulation in momentum space, but also of the SC pairing mechanism itself.  Finally, we also discuss the relevance of the Leggett mode in these two classes of materials for the non-linear optical response,\cite{cea_prb16} that has been shown to be experimentally accessible thanks to the use of intense multicycle THz pulses.\cite{shimano_science14} 

The structure of the paper is the following. In Sec. II we provide the derivation of the Raman response in the single-band case, to clarify the role of phase and density fluctuations. In Sec. III we derive the general form of the Raman response in a two-band superconductor, and comment on the simplified case of two equal band with opposite (hole/electron) character. In Sec. IV we show the evolution of the Raman response in the general two-band case from inter-band dominated to intra-band dominated pairing. The implications of our results, along with applications to non-linear optical spectroscopy, are discussed in Sec. V. The concluding remarks are reported in Sec. VI. Finally, Appendix A and B contain the technical details needed for the derivation of the Raman response in the single and two-band case, respectively. 

\section{Screening and gauge invariance in the single-band case}

To clarify the role of charge and SC phase fluctuations for the screening of the $A_{1g}$ Raman response we first outline the derivation of the Raman response for a single-band superconductor by means of the effective-action formalism. By introducing the Raman density operator $ \Phi_R(\mathbf{q})\equiv
\sum_{\mathbf{k}\sigma}\gamma(\mathbf{k})c^\dagger_{\mathbf{k}-\mathbf{q}/2,\sigma}c_{\mathbf{k}+\mathbf{q}/2,\sigma}$, where $\gamma_\bk$ is the Raman vertex, the Raman response is $S_R=-\frac{1}{\pi}[1+n(\omega)]\chi_{RR}''(\bq=0,\omega)$, where $n(\omega)$ is the
Bose-Einstein distribution and $\chi_{RR}(q)$ (with $q=(i\omega_n,\bq)$) is the Raman susceptibility after analytical continuation $i\omega_n\ra \omega+i\delta$. To derive $\chi_{RR}$ we will take advantage of the effective-action formalism, as detailed in Appendix A. We start from a microscopic fermionic model including the pairing $U$ and the Coulomb interaction $V(\bq)$, plus an external source field $\rho_R$ that is coupled to the fermionic Raman operator. As usual, one can decouple  the interacting terms  by means of the Hubbard-Stratonvich bosonic fields representing the collective fermionic degrees of freedom.\cite{nagaosa,depalo_prb99,randeria_prb00,benfatto_prb04}
After integrating out the fermions one is then left with an action expressed in terms of the relevant bosonic variables, i.e. 
 the SC amplitude, the SC  phase $\theta$, the electron density $\rho$ and  the Raman field $\rho_R$. The amplitude sector is as usual\cite{cea_cdw_prb14,cea_prl15} decoupled from the density/phase sector so one can write the effective action in the long-wavelength limit as:
\bea
\lb{sapprox}
S_{FL}&=&\frac{1}{2}\sum_q\left\{ |\rho_R(q)|^2\chi^0_{RR}(q)+\right.\nn\\
&+&\left.
2i\rho_R(-q)\chi_{R\rho}(-q)\left[\rho(q)+i\omega_n\theta(q)/2\right]+\right.\nn\\
&+&\left. \left(\frac{1}{V_\bq} - \chi_{\rho\rho}(q) \right) |\rho(q)|^2+\right.\nn\\
&+&\left. \frac{1}{4}\left(-\chi_{\rho\rho}\omega_n^2+D_s\bq^2\right)|\theta(q)|^2 +\right.\nn\\
&-&\left.  \chi_{\rho\rho}(q) \rho(-q) i\omega_n \theta(q) \right\}.
\eea
where $D_s$ denotes the superfluid stiffness, and we introduced the (bare) Raman-Raman ($\chi^0_{RR}$), the Raman-density ($\chi_{R\rho}$) and the density-density ($\chi_{\rho\rho}$) correlation  functions, given at $\bq=0$ by:
\bea
\lb{chizrr}
\chi^0_{RR}(i\omega_n)&=&-\sum_\bk \gamma_\bk^2 F_\bk(i\omega_n)\\
\lb{chirrho}
\chi_{R\rho}(i\omega_n)&=&- \sum_\bk \gamma_\bk F_\bk(i\omega_n)\\
\lb{chirhorho}
\chi_{\rho\rho}(i\omega_n) &=& -\sum_\bk  F_\bk(i\omega_n)
\eea
where
\be
\lb{fk}
F_\bk(i\omega_n)= 4\Delta^2
\frac{\tanh(E_\mathbf{k}/2T)}{E_\mathbf{k}\left[4E_\mathbf{k}^2-(i\omega_n)^2\right]},
\ee
and $E_\bk=\sqrt{\xi_\bk^2+\Delta^2}$, $\Delta$ being the SC gap.  
Finally, the full Raman susceptibility $\chi_{RR}$ can be computed from Eq.\ \pref{sapprox} by functional derivative with respect to the $\rho_R$ field, i.e.:
\begin{equation}\lb{chirr}
\chi_{RR}(q)=\left[\frac{\delta^2 {S}_{FL}}{\delta \rho_R(-q)\delta \rho_R(q)}\right]_{\rho_R=0}\quad.
\end{equation}
As one immediately sees from Eq.\ \pref{chirrho} for a non-symmetric Raman channel, such that $\sum_\bk \gamma_\bk=0$, the coupling $\chi_{R\rho}$ to the density and phase fluctuations vanishes, and the Raman response coincides with the bare one, Eq.\ \pref{chizrr}. On the other hand, in the $A_{1g}$ channel $\chi_{R\rho}\neq 0$ and one must add the effect of charge/density modes, described by the third and fourth line of Eq.\ \pref{sapprox}.\cite{depalo_prb99,randeria_prb00,benfatto_prb04,cea_prl15} In particular, the phase mode has a sound-like dispersion, that is converted in a plasmon-like mode when the density fluctuations are integrated out. 
Notice that while in the usual diagrammatic language the density mode is  included via RPA-like corrections and the phase mode via vertex-like corrections,\cite{hackl_review,klein_prb84} in the effective-action formalism they are both included by Gaussian integration of  $\rho,\theta$ in Eq.\ \pref{sapprox}. Moreover, they explicitly appear coupled to the Raman response by the same mixed susceptibility $\chi_{R\rho}$, showing that they must be always treated on the same footing. The final result for the Raman response function can be derived in a straightforward way in the limit $\bq=0$. Indeed, in this case Eq.\ \pref{sapprox} can be recast as:
\bea
\lb{sapprox2}
S_{FL}&=&\frac{1}{2}\sum_q\left\{ |\rho_R|^2\chi^0_{RR}+2i\rho_R\chi_{R\rho}\left[\rho+i\omega_n \theta/{2}\right]+\right.\nn\\
&+&\left. \frac{1}{V_\bq} |\rho|^2-
\chi_{\rho\rho}|\rho+i\omega_n{\theta}/{2}|^2 \right\},
\eea
making explicit the dependence only on the gauge-invariant combination $\rho+i\omega_n{\theta}/{2}$. Since $1/V_\bq\ra 0$ as $\bq\ra 0$ one can then shift $ \rho+i\omega_n{\theta}/{2}\ra\rho$ so that only the coupling of the Raman density to the density fluctuations appears explicitly. Then the integration of $\rho$ is straightforward and leads to the well-known result\cite{hackl_review,hackl_prl94,deveraux_prb95}
\be
\lb{chigi}
\chi_{RR}=\chi^0_{RR}-\frac{\chi_{R\rho}^2}{\chi_{\rho\rho}}.
\ee
For almost parabolic bands, where $\gamma_\bk\simeq 1/m$, one sees from Eq.s\ \pref{chizrr}-\pref{chirrho} that b0 th $\chi_{RR}^0$ and $\chi_{R\rho}$ are proportional to $\chi_{\rho\rho}$. In this condition 
Eq.\ \pref{chigi} guarantees that $\chi_{RR}=0$, as expected since the Raman response becomes proportional to the dynamical charge susceptibility, that must vanish at long wavelength. It is worth stressing that the vanishing of the dynamical charge susceptibility at $\bq=0$ is not a consequence itself of the presence of Coulomb interactions, but it is generally expected as a consequence of charge conservation.\cite{deveraux_prb95} Indeed, the charge susceptibility controls the charge redistribution in the presence on an external potential. Due to charge conservation, changing the charge density in one place can only be done by redistributing it, but this cannot be achieve with an uniform potential. On more general ground, the violation of this requirement for the BCS response function $\chi_{\rho\rho}$ in Eq.\ \pref{chirhorho} can be ascribed to the fact that in general  the BCS approximation is not gauge invariant,\cite{schrieffer} due to the fact that it accounts for the modification of the quasiparticle response in the SC state, but it does not include the effects of the SC collective modes. Only adding the contribution of the phase and density degrees of freedom one can restore the gauge invariance of the charge susceptibility, and in general of all the electromagnetic response functions\cite{depalo_prb99,randeria_prb00,benfatto_prb04}. 

According to the above discussion, the result \pref{chigi} must be independent on the presence of long-range Coulomb forces, and it must hold also when the interaction in the charge sector is short-ranged. This can be understood again from Eq.\ \pref{sapprox} by replacing $1/V_\bq$ with a generic short-range interaction $1/V$ (with $V\gtrless0$ for repulsive/attractive interaction). In this case the $|\rho|^2$ term in Eq.\ \pref{sapprox2} is finite, and one must integrate out {\em both} the density and the phase field. Indeed, after integrating out only $\rho$ one obtains (see Appendix A) for the Raman response $\tilde\chi_{RR}$ the result
\be
\lb{chisr}
\tilde\chi_{RR}=\chi^0_{RR}-\frac{\chi_{R\rho}^2}{\chi_{\rho\rho}-1/V},
\ee
being finite also when $\gamma_\bk=const$. On the other hand, by adding also the contribution of phase modes one immediately finds back the gauge-invariant result \pref{chigi}. This example clarifies that in the computation of the Raman response in the symmetric $A_{1g}$ channel the coupling to density and phase fluctuations must be treated on the same footing. On this respect, the approach used in the present work, based on the construction of the effective action including all the collective fluctuations coupled to the Raman response, is completely equivalent to the  diagrammatic derivation of the vertex corrections discussed e.g. in Ref.\ [\onlinecite{klein_prb84,hackl_review}], but with one additional advantage. Indeed, it allows one to recognize immediately that the vertex corrections in the particle-particle channel account for the fluctuations of the SC phase.  Since the phase is conjugate to the density, their effect must be always included along with RPA-like corrections in the particle-hole channel, which  account for density fluctuations.
While for the single-band case the presence of long-range Coulomb forces allows one to gauge away the phase mode,\cite{deveraux_prb95} making apparently its presence irrelevant, for a multiband system additional care should be used, since several phase modes appear. As we shall see in the next section, this crucial fact explains the difference between the results discussed so far in the literature within the context of MgB2\cite{blumberg_prl07,klein_prb10} or FeSC \cite{boyd_prb09,mazin_prb10,blumberg_prb10} for a multiband model with hole and electron pockets.

\section{Raman response in the two-band model}

The procedure discussed in the previous Section can be easily extended to a generic SC  two-band model, as detailed in the Appendix B. The microscopic starting point is the Hamiltonian:
\bea
\lb{h2b}
H&=&\sum_{\bk,\s,i} \xi_\bk^i c^{i,\dagger}_{\bk,\s}c^i_{\bk,\s} +H_P+H_C\\
\lb{hpair}
H_P&=&-\sum_{i,j,\bq}U_{ij}\Phi^{i,\dagger}_\D(\bq)\Phi^j_\Delta(\bq)\\
\lb{hc}
H_C&=&\sum_{\bq}V(\bq)\Phi^{\dagger}_\rho(\bq)\Phi_\rho(\bq)
\eea
where $i,j$ are the band indexes, $\Phi^i_\Delta(\bq)= \sum_\bk c^i_{-\bk+\bq/2,\down}c^{i}_{\bk+\bq/2,\up}$ and  $\Phi_\rho(\bq)= \sum_{\bk,i,\s} c^{i,\dagger}_{\bk-\bq/2,\s}c^i_{\bk+\bq/2,\s}$ are the pairing and density operators, respectively, $V(\bq)$ is the Coulomb potential and $\hat U\equiv U_{ij}$ is the matrix of the SC couplings.  
Notice that the interaction \pref{hpair} always assumes pairing of carriers within the same band, with opposite momenta at ${\bf q} = 0$. In addition, the pairing mechanism is
intra-band dominated when $\mathrm{det} \hat U= U_{11}U_{22}-U^2_{12}>0$,  while it is inter-band dominated when $\mathrm{det} \hat U<0$. 
The derivation of the Raman response follows the same strategy outlined for the single-band case. In particular by adding to the Hamiltonian \pref{h2b} a source field $\rho_R$ coupled to the total Raman density operator $ \Phi_R(\mathbf{q})=\sum_{i,\mathbf{k}\sigma}\gamma^i_{\mathbf{k}}c^{i,\dagger}_{\mathbf{k}-\mathbf{q}/2,\sigma}c^i_{\mathbf{k}+\mathbf{q}/2,\sigma}$ one can derive the effective action in terms of all the relevant collective modes coupled to the Raman density, equivalent to Eq.\  \pref{sapprox} above. As it has been discussed in Ref.\ \cite{marciani_prb13}, some special care has to be used to implement the Hubbard-Stratonovich transformation in the case of inter-band dominated pairing, i.e. $\mathrm{det} \hat U<0$. The technical details of the derivation are given in Appendix B, while we limit here the discuss to the main results and their physical implications.

\begin{figure}[htb]
\includegraphics[scale=0.3, angle=0,clip=]{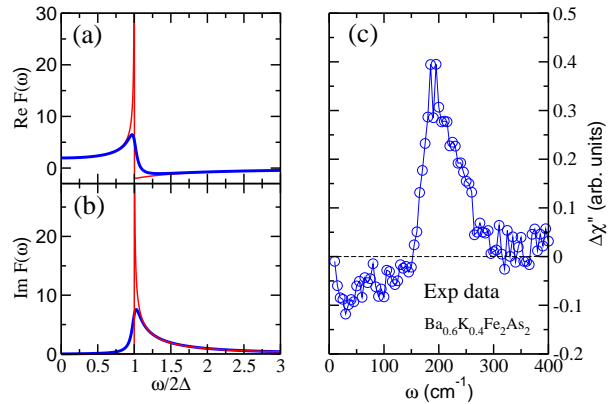}
\caption{(color online) (a) Real part and (b) imaginary part of the function $F(\omega)$ of Eq.\ \pref{fn} computed with $\delta\ra 0$ (red thin lines) and $\delta=0.01\Delta_0$ (blue thick lines).  According to Eq.\ \pref{chizrr} the $\mathrm{Im} F(\omega)$ is proportional to the unscreened (bare)  Raman response for a single-band superconductor. Its square-root divergence at $\omega=2\Delta$ signals the proliferation of the Cooper pairs break apart by the electromagnetic field. (c) Typical $A_{1g}$ spectrum for FeSC, as taken from Ref.\ \cite{hackl_prx14}. The strong enhancement of the signal at $\omega\simeq 150$ cm$^{-1}$ resembles the behavior of the unscreened Raman response, as it would be predicted by Eq.\ \pref{chiwrong}.}
\label{fig-F}
\end{figure}

For the multiband case the various susceptibilities \pref{chizrr}-\pref{chirhorho} depend now on the band index $i$ via both the Raman vertexes $\gamma_\bk^i$ and the functions $F^i_\bk$, computed on each band with dispersion $\xi_\bk^i$ and SC gap $\Delta_i$.  As it is shown in Eq.\ \pref{sfl2} of Appendix B, the Raman density $\rho_R$ is coupled both to the electron density and to the SC phase fluctuations in each band. By retaining {\em only} the coupling to the density fluctuations and integrating them out one recovers again the equivalent of the result \pref{chigi}, that now reads:
\be
\lb{chingi}
\chi_{RR}^D=\sum_i  \chi_{R_iR_i}^{0}-\frac{\left(\sum_i \chi_{R_i\rho_i}\right)^2}{\sum_i\chi_{\rho_i\rho_i}}.
\ee
This result, and its extension to more bands, has been used so far to interpret the experimental data in pnictides\cite{boyd_prb09,mazin_prb10,blumberg_prb10,hackl_prx14}.  To get a deeper insight into  the behavior of the expression \pref{chingi} let us first focus on the simplified case of one hole and one electron band, with parabolic energy dispersions, an approximation that can be good for FeSC. 
In the symmetric $A_{1g}$ channel the Raman vertex depends only on the electronic dispersion, so one has  $\gamma_\bk^i=1/m_i\equiv \gamma_i$, with $\gamma_1<0$ (hole band) and $\gamma_2>0$ (electron band). Even though the real band structure of FeSC has more than two bands, here we just model the main effect of having hole pockets at $\G$, and electron pockets at $X/Y$ (or at M in the 2Fe unit cell notation).
The Eq.\ \pref{chingi} can then be written\cite{blumberg_prb10} as:
\be
\lb{chid}
\chi^D_{RR}=-(\gamma_1-\gamma_2)^2\frac{N_1F_1N_2F_2}{N_1F_1+N_2F_2}
\ee
where $N_i$ is the density of states (DOS) in each band  and $F_i$ is the function obtained by integration over momenta in Eq.\ \pref{chirhorho}, i.e.
\be
\lb{fn}
F_i(i\omega_n)=4\Delta_i^2\int_{-\omega_D}^{\omega_D}\,d\xi\frac{\tanh\left[E_i(\xi)/2T\right]}{E_i(\xi)\left[4E_i(\xi)^2-(i\omega_n)^2\right]},
\ee
where $E_i(\xi)=\sqrt{\xi^2+\Delta_i^2}$ and $\omega_D$ is a typical cut-off for the SC interactions. In the limit where $\omega_D\gg \Delta_i$ Eq.\ \pref{fn} admits an analytical expression at $T=0$:
\bea
\mathrm {Re} F_i(\omega)&=&\frac{2\Theta(2\Delta_i-\omega)}{x_n\sqrt{1-x_i^2}}\arctan \frac{x_i}{\sqrt{1-x_i^2}}+\nn\\
\lb{ref}
&-&\frac{\Theta(\omega-2\Delta_i)}{x_n\sqrt{x_i^2-1}}\ln \frac{x_i+\sqrt{x_i^2-1}}{x_i-\sqrt{x_i^2-1}},\\
\lb{imf}
\mathrm {Im} F_i(\omega)&=&\frac{\Theta(\omega-2\Delta_i)\pi}{x_i\sqrt{x_i^2-1}}, \quad x_i=\frac{\omega}{2\Delta_i}
\eea
The real and imaginary parts of the function $F_i(\omega)$ are shown in Fig.\ \ref{fig-F}. According to Eq.\ \pref{chizrr}, the bare Raman response is proportional to $\mathrm{Im} F(\omega)$. Its square-root divergence at $\omega=2\Delta_i$ signals the proliferation of Cooper-pairs above this threshold. The behavior of the expression \pref{chid} is similar. 
In particular, when the two bands are equal $\gamma_1=-\gamma_2=\gamma$ and the two gaps coincide  $\Delta_1=\Delta_2=\Delta$ one immediately sees that Eq.\ \pref{chid} reduces to the {\em unscreened} single-band Raman response\cite{boyd_prb09,blumberg_prb10},
\be
\lb{chiwrong}
\chi^D_{RR}=-2\gamma^2 N F,
\ee
as it is evident also from Eq.\ \pref{chingi} due to the complete cancellation in this case of the term $\sum_i \chi_{R_i\rho_i}=0$ responsible for the screening. According to Eq.\ \pref{chiwrong} the Raman response should appear as the single-band unscreened case, see Fig.\ \ref{fig-F}b, in apparent agreement with the experiments in FeSC\cite{hackl_prb09,hackl_prl13,hackl_prx14,blumberg_prb16}, see e.g. the data reported in Fig.\ \ref{fig-F}c. Indeed, several experiments have shown so far that in FeSC the Raman response in the $A_{1g}$ channel is as large as in the other non-symmetric channels, with a shape that resembles the usual pair-breaking peak at the largest gap.

Despite its apparent agreement with experimental data in FeSC, the result \pref{chingi}  turns out to be in general incorrect. As we explained above, in the single-band case the coupling to the phase and density modes appears only via the gauge-invariant combination $\rho+i\omega_n\theta$. However, in the multiband case two different phase/density modes appear,\cite{leggett,sharapov_epjb02} the Goldstone mode $\theta_G=\theta_1+\theta_2$ and the Leggett one $\theta_L=\theta_1-\theta_2$.  Analogously to the the single-band case the Goldstone mode can be gauged away, and its contribution is already included in the result \pref{chingi} obtained by integrating out the total density, see Appendix B. However, the coupling to  the Leggett mode cannot be gauged away and its contribution must be added to Eq.\  \pref{chingi}. When this is correctly taken into account one finds that the final result is written in general as:
\begin{widetext}
\be
\lb{chitrue}
\chi_{RR}=\left(  \chi^0_{R_1R_1}+  \chi^0_{R_2R_2}  \right)-\frac{ (i\omega_n)^2\left(	\chi_{\rho_1\rho_1}\chi_{\rho_2R_2}^2+\chi_{\rho_2\rho_2}\chi_{\rho_1R_1}^2\right)+\kappa\left(\chi_{\rho_1R_1}+\chi_{\rho_2R_2}\right)^2  }{(i\omega_n)^2\chi_{\rho_1\rho_1}\chi_{\rho_2\rho_2}+\kappa\left(\chi_{\rho_1\rho_1}+\chi_{\rho_2\rho_2}\right)}
\ee
\end{widetext}
where
\be
\lb{kappa}
\kappa=\frac{8\Delta_1\Delta_2 U_{12}}{U_{11}U_{22}-U^2_{12}}
\ee
is a positive or negative constant depending on the nature of the pairing. The quantity on the numerator, $\Delta_1\Delta_2U_{12}$, is always positive irrespectively on the sign of $U_{12}$. Indeed, for inter-band attraction ($U_{12}> 0$) the gaps have the same sign, while for inter-band repulsion ($U_{12}<0$) the gaps must have opposite sign. On the other hand the quantity on the denominator, i.e. $\mathrm{det} \hat U=U_{11}U_{22}-U^2_{12}$, depends on the nature of the pairing, being
positive for intra-band dominated pairing, where $\mathrm{det} \hat U>0$ and $\kappa>0$, and negative in the opposite case of interband
dominated pairing, where $\mathrm{det} \hat U<0$ and $\kappa<0$. Once more, in the case of parabolic bands the expression \pref{chitrue} simplifies leading to:
\be
\lb{chirrapp}
\chi_{RR}=(\gamma_1-\gamma_2)^2\frac{\kappa }{(i\omega_n)^2-F_L(i\omega_n)}
\ee
where we introduced the function $F_L$:
\be
\lb{fl}
F_L(i\omega_n)=\kappa \frac{(N_1F_1+N_2F_2)}{N_1F_1N_2F_2}.
\ee
Eq.\ \pref{chirrapp} has been derived by  means of a standard diagrammatic implementation of vertex corrections in Ref.\ \cite{blumberg_prl07,klein_prb10}, and it has been used to interpret the experiments in MgB$_2$. Its physical interpretation is straightforward: while for equal bands having same character $\gamma=1=\gamma_2$ the $A_{1g}$ Raman response vanishes because of charge conservation, when the two bands have opposite character, i.e. $\gamma_1=-\gamma_2$, the Raman density scales as the relative density fluctuations $\rho_1-\rho_2$. As such, it couples to the relative phase Leggett mode $\theta_L=\theta_1-\theta_2$, whose energy $\omega_L$ is identified\cite{leggett,sharapov_epjb02}, by the vanishing of the denominator of Eq.\ \pref{chirrapp}: 
\be
\lb{eqleg}
\omega^2_L-F_L(\omega_L)=0.
\ee
When the interband coupling $U_{12}$ is small the solution of Eq.\ \pref{eqleg} can be found by taking the limit $F_L(\omega\ra 0)$. Since $F_i(0)\simeq2$, see Eq.\ \pref{fn}, one sees that the energy $\omega_L^2$ of the Leggett mode is
\be
\lb{omegal}
\omega_L^2= F_L(0)\simeq\kappa \frac{N_1+N_2}{2N_1N_2}=\frac{4\Delta_1\Delta_2 U_{12}}{\mathrm{det} \hat U}\frac{N_1+N_2}{N_1N_2}
\ee
in agreement with the result found by Leggett\cite{leggett} for intra-band dominated pairing. For larger interband coupling $\omega_L$ is found numerically from Eq.\ \pref{eqleg}, but it always lies below the largest gap (see next Section), giving rise to a sharp resonance in the $A_{1g}$ channel.

\begin{figure}[htb]
\includegraphics[scale=0.4, angle=0,clip=]{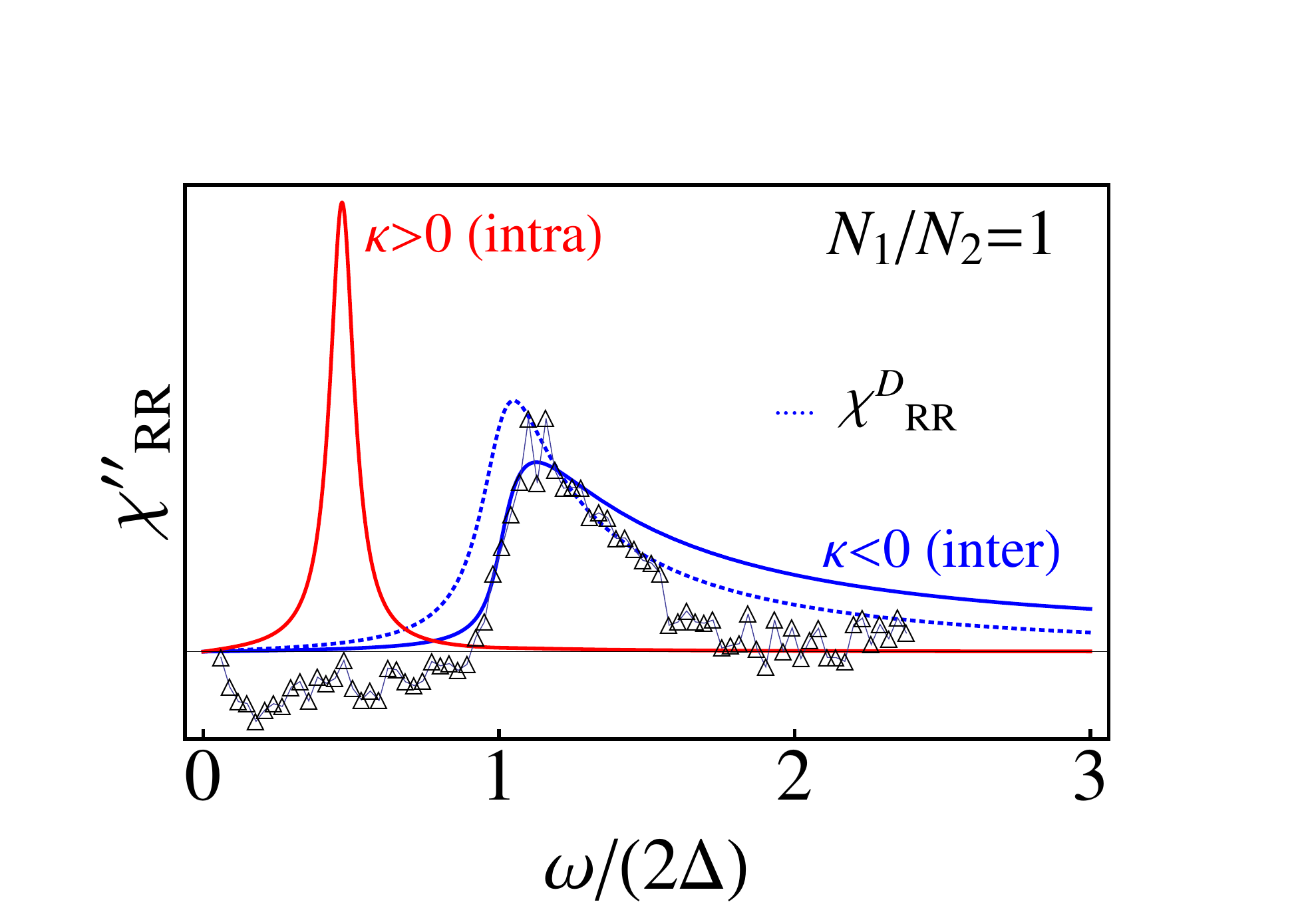}
\caption{(color online) Raman response for the case of two bands having same density of states and same SC gap, but opposite character, $\gamma_1=-\gamma_2=\gamma$. The solid line is the full gauge-invariant result \pref{chicorr} for intraband ($\kappa>0$, red line) and interband ($\kappa<0$, blue line) dominated pairing. The dashed line represent the result \pref{chiwrong}, obtained by including only the contribution of density fluctuations. Here a residual damping $\delta=0.01\Delta$ has been used in the analytical continuation of Eq\ \pref{fn}, that smears out the divergence of the function $F(\omega)$ at $\omega=2\Delta$. The symbols represent the experimental data of Fig.\ \ref{fig-F}, taken from Ref.\ \cite{hackl_prx14}. As one can see, they are consistent with Eq.\ \pref{chicorr} in the case of dominant interband pairing. }
\label{fig-comp}
\end{figure}

\begin{figure*}[t]
\includegraphics[scale=0.4, angle=0,clip=]{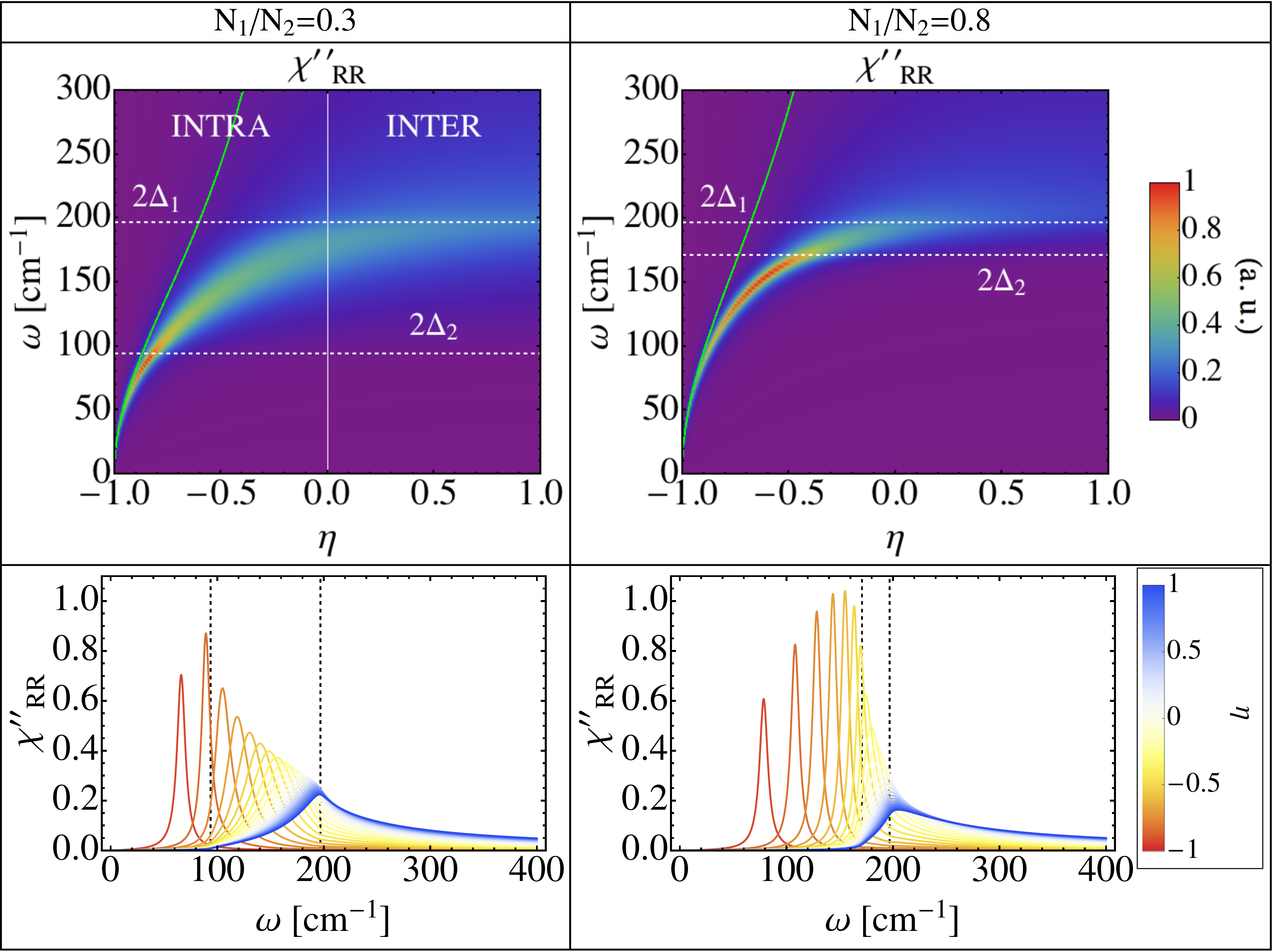}
\caption{(color online) Top panels: colour plot of the Raman response in the $A_{1g}$ channel according to Eq.\ \pref{chirrapp} as a function of the interband coupling constant, defined by the dimensionless quantity \pref{defg}. Here $\eta=-1$ is the the intra-band only case, while $\eta=1$ is the inter-band only case. The dashed lines indicate the (absolute) values of twice the gaps. The red green line denotes the analytical expression \pref{omegal} for the value of the Leggett mode in the weak-coupling regime. Bottom panels: cuts of the frequency-dependent Raman response for selected values of the couplings. As soon as one enters the inter-band dominated regime the Leggett resonance appears as a broad feature peaked slightly above $2\Delta_1$, that resembles the usual pair-breaking peak of the unscreened, single-band Raman response. }
\label{fig-raman}
\end{figure*}

For a system with inter-band dominated pairing is $\kappa<0$, so that the expression \pref{omegal} does not admit a solution, intended as a sharp resonance below the quasiparticle threshold.  Even though this condition rules out the existence of a true Leggett mode\cite{maiti_prb13,marciani_prb13}, nonetheless it does not rule out the unavoidable coupling of the Raman response to relative phase fluctuations $\theta_L=\theta_1-\theta_2$. As a consequence, the correct expression, Eq.s\ \pref{chitrue} and \pref{chirrapp}, for the Raman response in the $A_{1g}$ channel must be used irrespectively of the intra/inter-band nature of the pairing. The crossover from intra-band dominated to inter-band dominated regime for the Raman response can be simply understood resorting to the simplified case of two bands with equal gap and opposite character, $\gamma_1=-\gamma_2=\gamma$. In this case Eq.\ \pref{chirrapp} reduces to:
\bea
\chi_{RR}(\omega)&=&\frac{4\kappa \gamma^2}{(\omega+i\delta)^2-2\kappa/(NF(\omega))}=\nn\\
\lb{chicorr}
&=&\frac{-2\kappa \gamma^2NF(\omega)}{1-NF(\omega)(\omega+i\delta)^2/2\kappa}
\eea
that has to be contrasted to the result \pref{chiwrong}. From Eq.\ \pref{ref} one immediately sees that $\mathrm{Re}F(\omega)$ diverges as  $\omega\ra(2\Delta)^-$, where it also changes sign from positive to negative. In this situation the denominator of Eq.\ \pref{chicorr} vanishes at $\omega<2\Delta$ when $\kappa>0$. Since in this regime $\mathrm{Im}F=0$, see Eq.\ \pref{imf}, the resulting mode is sharp since it is undamped by quasiparticles, see Fig.\ \ref{fig-comp}. On the other hand when $\kappa<0$ the real part of the denominator of Eq.\ \pref{chicorr} can only vanish at $\omega>2\Delta$, where $\mathrm{Re}F$ becomes negative. However, since at $\omega>2\Delta$ also $\mathrm{Im}F $ starts to develop this resonance is always strongly overdamped, and $\chi"_{RR}$ from Eq.\ \pref{chicorr} is dominated by the imaginary part of the numerator. This is the reason why the $A_{1g}$ channel displays a resonance right above the gap that can be qualitatively similar to the unscreened result obtained with the wrong expression \pref{chiwrong}, especially when a small residual damping is taken into account, see Fig.\ \ref{fig-comp}. More importantly, as shown in 
Fig.\ \ref{fig-comp} this result is in good agreement with experimental data in FeSC, even within the simplified case of two equal bands.  As we shall see in the next Section, by considering a more general multiband model  with different DOS and gap values in the two bands the qualitative differences  between the two results \pref{chirrapp} and \pref{chid} become more evident.

\section{Leggett resonance from inter-band to intra-band pairing}

To analyze the general evolution of the Leggett-mode resonance from intra-band dominated to inter-band dominated coupling we study the case of two parabolic bands with opposite character, where the expression \pref{chirrapp} holds. In the limit where only $U_{12}\neq 0$, i.e. when pairing is provided uniquely by interband interactions, one can easily sees that the gaps must satisfy $\Delta_1/\Delta_2=sign(U_{12})N_2/N_1$ at $T=0$\cite{benfatto_prb08}. To mimic the case of FeSC, where an interband repulsion is expected, we then take $U_{12}<0$. By fixing the value of $N_1/N_2$ we can then vary the SC coupling $U_{ij}$ from $det\hat U>0$ to $det\hat U<0$, by retaining the same values of $\Delta_1>\Delta_2$.  If we define the dimensionless quantity:
\be
\lb{defg}
\eta=\frac{|U_{12}|-\sqrt{U_{11}U_{22}}}{|U_{12}|+\sqrt{U_{11}U_{22}}}
\ee
one immediately sees that $\eta$ goes from -1 to +1 as the interband coupling increases, so that $\eta=-1$ is the case where $U_{12}=0$ while $\eta=1$ is the case where $U_{11}=U_{22}=0$. 

The full Raman response obtained from the expression \pref{chirrapp} is shown in Fig.\ \ref{fig-raman}. As one can see, in agreement with the simplified case of two equal bands, in the range $\eta<0$ the Leggett mode, given by the solution of Eq.\ \pref{eqleg}, identifies a sharp resonance below the largest gap, whose spectral weight is maximum as $\omega_L$ approaches the smallest gap. For very weak interband coupling $\omega_L\ra 0$ so it follows the analytical expression \pref{omegal}, see solid line in Fig.\ \pref{fig-raman}. On the other hand for larger interband coupling one cannot neglect the frequency dependence \pref{fl} of the $F_L(\omega)$ function appearing in Eq.\ \pref{eqleg}, reflecting the breaking of Cooper pairs at $\omega>2\Delta_2$, so that  $\omega_L$ deviates considerably from the low-energy limit \pref{omegal} and it is finally limited by the upper bound $2\Delta_1$ given by the largest gap. In a recent numerical analysis\cite{manske_natcomm16} of collective modes in a two-band superconductor, this effect has been attributed to the coupling between the Leggett mode and the amplitude modes, that is zero in the particle-hole symmetric $\bq=0$ limit considered here, but becomes finite in the case of finite external momentum considered in Ref.\ [\onlinecite{manske_natcomm16}]. Even though this coupling can modify the expression \pref{eqleg}, we believe that the softening of the Leggett mode with respect to the low-frequency result \pref{omegal} observed in Ref.\ \cite{manske_natcomm16} can be simply understood as a consequence of the interplay between the Leggett mode and the quasiparticle continuum, encoded in Eq.\ \pref{eqleg}, and shown in Fig.\ \ref{fig-raman}. Notice also that when $\eta\ra -1$, i.e. the interband coupling goes to zero $U_{12}\ra 0$, the signature of the Leggett mode in the Raman response disappears. This can be easily understood from Eq.\ \pref{chirrapp}, considering that $\kappa\propto U_{12}$ as $U_{12}\ra0$, see Eq.\ \pref{kappa}. As a consequence when  $U_{12}\ra 0$ the Leggett mode, given by Eq.\ \pref{omegal}, scales as $\omega_L\propto \sqrt{\kappa}$, so that the imaginary part of Eq.\ \pref{chirrapp} reads:
\bea
\chi"_{RR}&\simeq& \frac {(\gamma_1-\gamma_2)^2 \kappa}{2\omega_L}\delta(\omega-\omega_L)\propto \nn\\
&\propto& (\gamma_1-\gamma_2)^2 \sqrt{\kappa}\delta(\omega-\omega_L)\ra 0, \quad \kappa\ra 0
\eea
This result is again consistent with the fact that when the bands are decoupled the Raman response in the $A_{1g}$ channel can only probe the total density fluctuations, that must vanish by gauge invariance in the long-wavelength limit.

In the regime of interband-dominated coupling, i.e.  $\eta>0$,  Eq.\ \pref{eqleg} cannot have a solution for $\omega<2\Delta_1$. Indeed, by closer inspection of Eq.\ \pref{fl} one sees that $\mathrm{Re} F_L(\omega)$  becomes negative only at 
 $\omega>2\Delta_1$ where both $\mathrm{Re} F_1$ and $\mathrm{Re}F_2$ are negative, compensating the negative sign of the prefactor $\kappa<0$. However, since at $\omega>2\Delta_1$ also the two imaginary parts of $F_1$ and $F_2$ are different from zero, the overall spectral function has always a maximum at $\omega\simeq 2\Delta_1$, i.e. at the largest of the two gaps, with an overall intensity quite smaller than in the intra-band dominated regime. When the two gaps have similar values, see right panels in Fig.\ \pref{fig-raman}, the resulting Raman response resembles qualitatively the case of identical bands with opposite character shown in Fig.\ \ref{fig-comp} above.

\begin{figure}[htb]
\includegraphics[scale=0.4, angle=0,clip=]{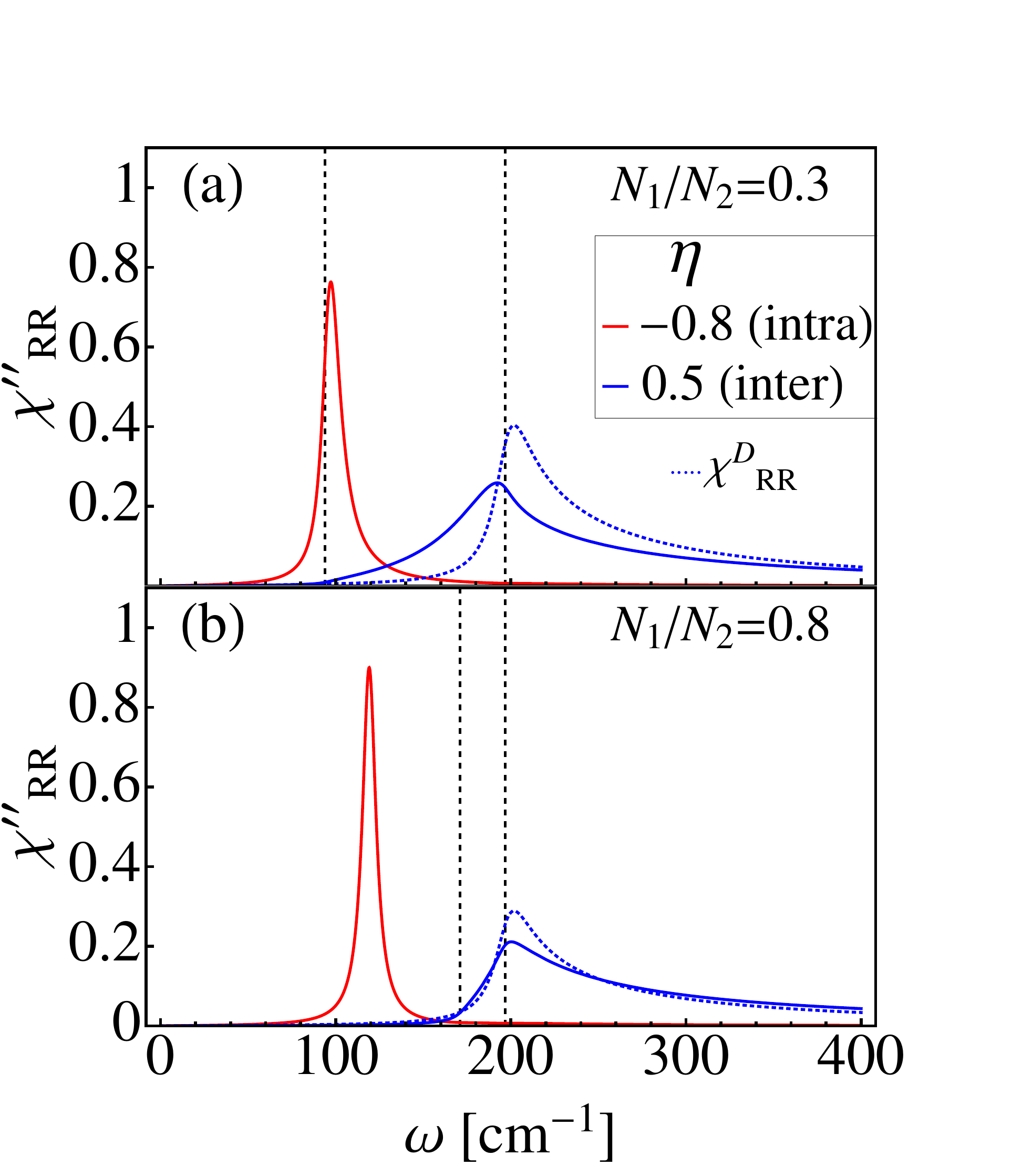}
\caption{(color online) Raman response in the $A_{1g}$ for intraband dominated pairing ($\eta<0$, red line) and interband dominated pairing ($\eta>0$, blue line) for two different values of the ratio $N_1/N_2$. As in Fig.\ \ref{fig-raman} the two gap values $\Delta_2<\Delta_1$ are kept fixed, and identified by the vertical dashed lines. The dotted blue line represents $\chi_{RR}^D$ from Eq.\ \pref{chid} for the same values of the band parameters. As one can see, in the case (a) where the two gaps are rather different $\chi_{RR}^D$  differs qualitatively by the gauge-invariant result \pref{chirrapp} for both value of $\eta$. On the other hand, in the case (b) one observes an accidental similarity between $\chi_{RR}^D$  and the correct expression, due to the fact that in this case the Leggett resonance itself resembles an unscreened Raman response.}
\label{fig-final}
\end{figure}

In Fig.\ \ref{fig-final} we summarize the results for the $A_{1g}$ Raman response for the model with one hole-like and one electron-like band, as a function of the interband coupling. For the sake of completeness we also show with dashed lines the result for $\chi_{RR}^D$, Eq.s \pref{chingi}-\pref{chid}, where only the contribution of the density modes is included. As one can see, in the case of inter-band dominated coupling ($\eta=-0.8$, red curves) the Eq.\ \pref{chingi} completely fails to recover the Leggett resonance. Indeed,  $\chi_{RR}^D$ is always peaked at the largest gap, the peak at the smallest gap being removed by the second term of Eq.\ \pref{chingi} which is not zero in this case. We notice that the case $N_1/N_2=0.3$ and $\eta=-0.8$, so that $\Delta_1\simeq 2\Delta_2$, exemplifies the situation for MgB$_2$,\cite{review_mgb2,sharapov_epjb02,klein_prb10} where the two gaps have rather different values and the Leggett mode is expected to lie between them, as indeed observed experimentally\cite{blumberg_prl07}. 

On the other hand, when the system has dominant interband coupling ($\eta=0.5$) the Raman response is always peaked at the largest gap, with a tail starting already at the smallest one, see  Fig.\ \ref{fig-final}a, being also in this case qualitatively different from the result \pref{chid}. Even though for similar gap values (panel b) the difference becomes less relevant, making the two results accidentally similar,  the physical mechanisms behind them are completely different. Indeed, while the expression \pref{chid} attributes the resonance in FeSC to a pair-breaking mechanism, made visible by lack of Coulomb screening,\cite{boyd_prb09,mazin_prb10,blumberg_prb10,hackl_prx14} 
the expression \pref{chicorr} always identifies the $A_{1g}$ resonance with a Leggett mode, whose nature in turn depends on the intra- vs inter-band character of the pairing. On this respect, the comparison with experimental data in FeSC suggests that the $A_{1g}$ Raman response in FeSC provides an indirect evidence on the {\em interband} nature of the pairing in these systems, supporting the proposal\cite{si_review16,hirschfeld_review16}   that pairing in FeSC is mediated by the exchange of spin fluctuations between hole and electron pockets. Indeed, such a pairing mechanism has a predominant inter-band character, so that $det \hat U<0$ and consequently also $\kappa<0$ in Eq.\ \pref{kappa}. This would explain the lack of a sharp sub-gap mode in the $A_{1g}$ channel of FeSC, and the observation of a sizeable signal peaked approximately at twice the SC gap estimated by other measurements.\cite{hackl_prb09,hackl_prl13,hackl_prx14,blumberg_prb16} We also notice that as far as the pairing mechanism is inter-band dominated this result is also robust with respect to the presence of accidental nodes\cite{si_review16,hirschfeld_review16} of the gaps in one of the
bands, even though in this case longer tails below twice the gap maxima could be expected.

The model \pref{h2b}-\pref{hc} provides a rather general description of the SC properties of a multiband system. However, additional interactions could be present, specific to a given system. For the $A_{1g}$ channel of FeSC it has been suggested \cite{chubukov_prb09}  that also short-range density interactions in the $\rho_1-\rho_2$ channel should be included. In this case, the Raman response in the $A_{1g}$ channel couples also to relative density fluctuations, whose integration can lead to a contribution analogous to Eq.\ \pref{chisr}, where now $V<0$. One can then easily understand that since $\chi_{\rho\rho}$ from Eq.\ \pref{chirhorho} is proportional to $F(\omega)$, the divergence of its real part at $\omega=2\Delta$ leads to a sharp sub-gap mode,  as shown in Ref.\ [\onlinecite{chubukov_prb09}]. This mechanism is somehow analogous to the one discussed in Ref.\ [\onlinecite{gallais_prl16}] for the $B_{1g}$ channel,\cite{note_raman} where the Raman response couples to nematic density fluctuations having  the same $B_{1g}$ symmetry, leading to a sub-gap resonance. This mechanism can be responsible for the subgap resonance observed in several FeSC in the 
$B_{1g}$ channel\cite{gallais_prl16,hackl_prl13,hackl_prx14,blumberg_prb16}, even though it has been also attributed to a 
Bardasis-Schrieffer mode,\cite{hackl_prl13,hackl_prx14} due to the presence of an additional sub-leading pairing channel. For what concerns the $A_{1g}$ channel the signatures observed in FeSC has been mainly attributed to an unscreened Raman signal\cite{boyd_prb09,mazin_prb10,blumberg_prb10,hackl_prx14}, even though recent data in 1111 NaFe$_{1-x}$Co$_x$As samples have been interpreted in terms of the sharp sub-gap resonance predicted in Ref.\ \cite{chubukov_prb09}. While this is an open possibility, one should notice that this interpretation is based mainly on the fact that the $A_{1g}$ resonance emerges below twice the largest gap, whose value is estimated by ARPES measurements on electronic pockets and the outer hole pocket.\cite{arpes1111} Indeed, its profile does not resemble a sharp mode, but it is similar to previous observations in 122 compounds.,\cite{hackl_prl13,hackl_prx14} In particular, by assuming that a lower gap opens also on the inner $\alpha$ hole pocket, that barely crosses the Fermi level in the normal state,  the profile of the $A_{1g}$ signal reported in Ref.\ \cite{blumberg_prb16} could be easily compared with the results of Fig.\ \ref{fig-final}a, obtained for gaps with marked  different values. This interpretation would allows one also to estimate the SC gap on the inner $\alpha$ pocket, that cannot be easily resolved by ARPES measurements.\cite{arpes1111}. 

\begin{figure*}[t]
\includegraphics[scale=0.4, angle=0,clip=]{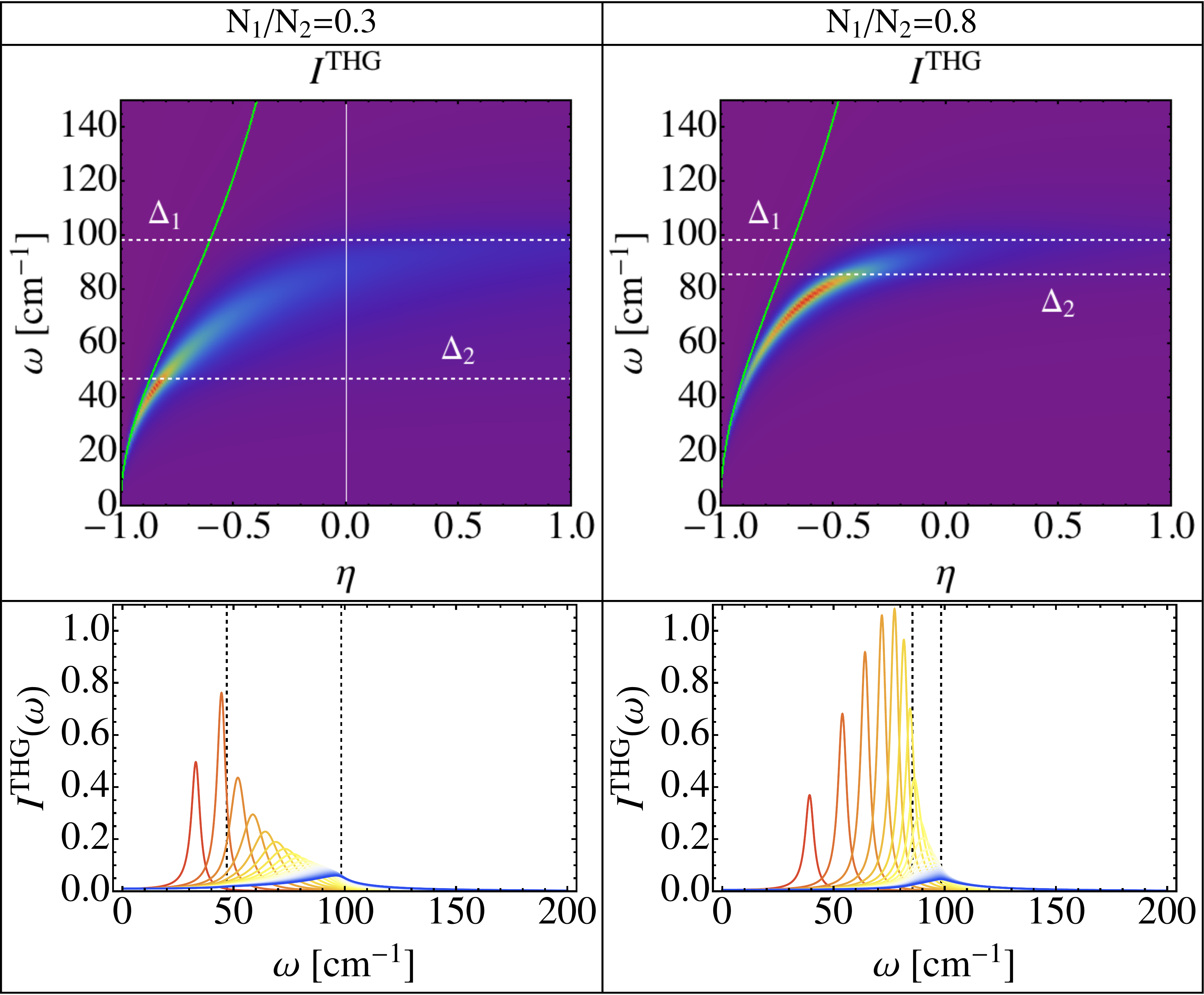}
\caption{(color online) Top panels: colour plot of the intensity of the THG according to Eq.\ \pref{thg} as a function of the interband coupling constant $\eta$, with the same parameters used in Fig.\ \ref{fig-raman}. Since $I^{THG}(\omega)$ is proportional to the Raman response computed at twice the light frequency $\chi_{RR}(2\omega)$, we marked with dashed lines indicate the (absolute) values of the gaps, and with the red green line $\omega_L/2$ as given by the analytical expression \pref{omegal}. Bottom panels: cuts of the THG intensity for selected values of the interband coupling. }
\label{fig-thg}
\end{figure*}

\section{Third-harmonic generation in non-linear optics}
Finally, we would like to briefly discuss the connection between the present results, derived in the context of the Raman response, and the non-linear optical response in the THz frequency range of a two-band superconductor. In the last few years the advances in the generation of intense multycicleTHz fields has shown that  non-linear optical effects become experimentally accessible. This has been clearly demonstrated in Ref. [\onlinecite{shimano_science14}] by the observation, in a BCS superconductor, of a component of the electromagnetic field oscillating three time faster than the incoming light. As it has been recently discussed in Ref.\ \cite{cea_prb16}, the third-harmonic generation (THG) can be understood by computing the equilibrium non-linear response, that turns out to measure lattice-modulated density correlations, in close  analogy with Raman spectroscopy. More specifically, one can see that the non-linear current $J^{NL}$ is given by
\be
\lb{jnl}
J^{NL}_\a(t)=-2e^4A_\a(t)\int\,dt'\sum_\b K_{\a\b}(t-t')A^2_\b(t'),
\ee
where $\bA$ is the e.m. gauge field, $\a,\b=x,y$ denote the spatial components and the response function $K_{\a\b}$ is given by
\be
\lb{kij}
K_{\a\b}(i\Omega_n)=\langle \rho_\a \rho_\b\rangle
\ee
with the operator $\rho_\a$ defined as:
\be
\rho_\a(\bq)=\sum_\bk (\pd^2 \e_\bk/\pd k^2_\a) c^\dagger_{\bk+\bq}c_\bk.
\ee
As a consequence, the non-linear response kernel $K_{\a\b}$ in Eq.\ \pref{kij} probes density fluctuations, with the inverse mass tensor $(\pd^2 \e_\bk/\pd k^2_\a)(\pd^2 \e_\bk/\pd k^2_\b)$ accounting for the relative direction of the incoming e.m. field $\bA$ with respect to the main crystallographic axes, in full analogy with the Raman response where the $\gamma_\bk$ vertex accounts for the polarization of the incoming and outgoing light. In the limit of parabolic hole/electron bands with mass $m$, that is the case considered here, $(\pd^2 \e^i_\bk/\pd k^2_\a)=\pm1/m\equiv \gamma_i$. In this case it is easy to see that the multiband non-linear kernel $K_{\a\b}=K$ is independent on the spatial indexes $\a\b$ and it coincides exactly with the multiband Raman response in the $A_{1g}$ channel computed so far.   By considering e.g. an incident  monocromatic field $\bA=A_0 \cos(\omega t)\hat x$ it is easy to show\cite{cea_prb16} from  Eq.\ \pref{jnl} that the non-linear current $J^{NL}_x$ has a component oscillating at $3\omega$, whose intensity  is defined as 
$I^{THG}(\omega)\propto \left|	\int\,dt J^{NL}_x(t)e^{3i\omega t} \right |^2$ and it is given by
\begin{equation}
\lb{thg}
	I^{THG}(\omega)=I_0e^8A_0^6\left|	K(2\omega)	\right|^2\equiv I_0e^8A_0^6\left|\chi_{RR}(2\omega)	\right|^2\quad,
\end{equation} 
with $I_0$ an overall scale factor, and $\chi_{RR}$ given by Eq.\ \pref{chirrapp}, valid in the case of parabolic bands. The corresponding evolution of the THG intensity is shown in Fig.\ \ref{fig-thg} for the same range of parameters of Fig.\ \ref{fig-raman}. Notice that in Eq.\ \pref{thg} it appears the modulus of the complex response function $\chi_{RR}(\omega)$, that differs from the Raman response that only probes $\chi"_{RR}$. Nonetheless, we still find in Eq.\ \ref{fig-thg} that for a fixed value of the interband coupling the non-linear response has a maximum when $2\omega=\omega_L$ matches the Leggett-mode frequency. On the other hand, as already observed in the case of the Raman response, the overall spectral weight of the Leggett resonance is rapidly suppressed at small interband coupling $\eta\ra -1$, and  it essentially disappears as soon as one enters the inter-band dominated regime $\eta>0$. As we mentioned above, the MgB$_2$ superconductor can be very well described by approximate parabolic bands: we then expect that {\em only} the Leggett mode contributes to the THG, in agreement with the experimental observation of the single Leggett resonance in Raman experiments\cite{blumberg_prl07}. Indeed, even though in the general lattice case\cite{cea_prb16} also the density fluctuations can give a resonant contribution at $2\omega=2\Delta_{1,2}$, corresponding to the first two terms of Eq.\ \pref{chitrue}, when $\gamma_i\simeq const$ in each band the only remaining resonance is the Leggett mode. On the other hand, the strong suppression of the Leggett resonance in the case of interband-dominated interactions suggests that in FeSC the observation of the Leggett resonance via non-linear THz optical spectroscopy is quite unlikely, so that only density-like resonances at $\omega=\Delta_{1,2}$, triggered by non-parabolic lattice structures, are possible. To quantify this effect one needs to resort to a specific lattice band dispersion, that is beyond the scope of this paper. However, we stress that also in the more general case of a lattice model the equivalence between the general $K_{\a\b}$ response function and the general Raman response function \pref{chitrue} still holds, provided that the $\gamma_\bk$ insertion in the Eq.s\ \pref{chizrr}-\pref{chirrho} are replaced by the derivatives of the dispersion in each band according to the general prescription \pref{kij}.

\section{Conclusions}

In the present work we used an effective-action formalism to derive the general expression for the  Raman response of a two-band superconductor. We have explicitly shown that even though in the usual diagrammatic approach the contribution of density and SC phase modes originate from different (RPA-like or vertex-corrections like) subset of diagrams, their contributions must be always treated on the same footing, in order to obtain the correct results. As an example, we discussed how in the single-band case the so-called notion of "Coulomb screening" in  the symmetric $A_{1g}$ Raman channel is somehow misleading. Indeed, the vanishing of the Raman response when the Raman density is proportional to the full density is a general consequence of charge conservation and gauge invariance, that can only be restored by adding the contribution of both charge and SC phase fluctuations. Nonetheless, it is also true that in the single-band case the presence of long-range Coulomb forces allows one to gauge away the phase mode, recovering the gauge-invariant Raman response in the $A_{1g}$ channel by adding only the RPA-like resummation of density fluctuations. 

In the multiband case the same procedure allows one to eliminate the SC Goldstone phase mode, but not the Leggett mode, that turns out to dominate the response of the $A_{1g}$ channel in the case of two bands having opposite (hole and electron) character. Interestingly, this result has been correctly pointed out in the context of MgB$_2$ superconductors,\cite{blumberg_prl07,klein_prb10}  but has been completely overlooked so far in the discussion of the Raman response of iron-based superconductors.\cite{boyd_prb09,mazin_prb10,blumberg_prb10,hackl_prx14} Indeed, by taking into account only the effect of charge fluctuations the main focus has been put so far on the lack of Coulomb screening when the two bands have opposite character. According to this interpretation the large $A_{1g}$ signal observed experimentally in FeSC\cite{hackl_prb09,hackl_prl13,hackl_prx14,blumberg_prb16} should be attributed to the unscreened pair-breaking peak. In this paper we explain why this result is formally not correct, and we show that also a Leggett resonance can account for the experimental data in FeSC, provided that the pairing has a dominant {\em interband} character. Indeed, in this case a true Leggett mode, intended as a sharp peak below the quasiparticle excitation threshold, cannot exist, as already discussed previously within the context of FeSC.\cite{maiti_prb13,marciani_prb13} Nonetheless, the Josephson-like fluctuations of the SC phases of the two condensates still identify a resonance pushed right above the largest gap, in close resemblance with Raman experiments in FeSC.

As we summarize in Fig.\ \ref{fig-final}, the full gauge-invariant result can be accidentally similar to an unscreened Raman pair-breaking peak when the pairing is inter-band dominated and two gaps are similar. However, this coincidence does not rule out the profound difference between the two physical mechanisms discussed here and in previous work on FeSC.\cite{boyd_prb09,mazin_prb10,blumberg_prb10,hackl_prx14} We notice also that the effects of charge and phase modes are not simply additive.  Indeed, after integrating out the relative phase modes the divergence of the unscreened response, obtained by considering only the density modes, is removed in favour of the Leggett resonance, that is the only one visible in the $A_{1g}$ channel for two bands having opposite character. As we discussed above this result is also relevant for non-linear optical spectroscopy in multiband superconductors, where the intensity of the so-called third-harmonic generation observed so far in single-band superconductors\cite{shimano_science14} is controlled by a response function analogous to the Rama one. In particular we expect that in MgB$2$, where the bands are approximately parabolic, only the Leggett mode contributes to the THG, while in FeSC the Leggett contribution appears too small to be detected. 

As we discussed in this manuscript, an interesting outcome of our results is the possibility to use Raman measurements to establish the nature (intra- vs inter-band) of the pairing mechanism in FeSC. So far, Raman results have been provided mainly for 122 systems, where quite a wide consensus already exists that spin fluctuations can provide an efficient mechanism for interband pairing between hole and electron pockets, leading in turn to a $s_\pm$ symmetry of the order parameter.\cite{si_review16,hirschfeld_review16} On the other end, the situation is more controversial for other systems, like LiFeAs or FeSe,\cite{buchner_natmat14,meingast_prl15,borisenko_prl10,vanderbrink_prl11} where alternative gap symmetries have been proposed, eventually compatible with intra-band pairing mechanisms, as provided e.g. by phonons or more unconventional orbital fluctuations.\cite{si_review16,hirschfeld_review16}  As a consequence, our results for the $A_{1g}$ Raman response pave the way to an alternative route to investigate the nature of the pairing interaction, that can be used to asses the relevant glue mechanism at play in FeSC.

\acknowledgements
We acknowledge useful discussions with T. B\"ohm, T. P. Deveraux, R. Hackl and P. Hirschfeld. We thank in particular T. B\"ohm and R. Hackl for a crucial exchange of ideas on the Raman response on iron-based superconductors, and for providing us with the experimental data shown in Fig.\ \ref{fig-comp}.  
This work has been supported  by Italian MIUR under projects FIRB-HybridNanoDev-RBFR1236VV, PRINRIDEIRON-2012X3YFZ2 and Premiali-2012 ABNANOTECH.

\begin{appendix}

\section{Derivation of the Raman response for the single-band case}
We model a generic single-band $s$-wave superconductor via the following Hamiltonian:
\bea\label{model}
H&=&\sum_{\bk,\s} \xi_\bk c^{\dagger}_{\bk,\s}c_{\bk,\s} +H_P+H_C\\
\label{modelpair}
H_P&=&-U\sum_{\bq}\Phi^{\dagger}_\D(\bq)\Phi_\Delta(\bq)\\
\label{modelc}
H_C&=&\sum_{\bq}V(\bq)\Phi^{\dagger}_\rho(\bq)\Phi_\rho(\bq)
\eea
where $\xi_\bk$ is the band dispersion, $\Phi_\Delta(\bq)= \sum_\bk c_{-\bk+\bq/2,\down}c_{\bk+\bq/2,\up}$ and $\Phi_\rho(\bq)= \sum_{\bk,\s} c^{\dagger}_{\bk-\bq/2,\s}c_{\bk+\bq/2,\s}$ are the pairing and density operators, respectively, $V(\bq)$ is the Coulomb potential and $U>0$ is the SC coupling.

To better describe the SC pairing introduced by $H_P$ it is useful to represent the fermions via the Nambu spinor $\Psi_\mathbf{k}=\begin{pmatrix}c_{\bk\uparrow}	,	c^\dagger_{-\bk\downarrow}	\end{pmatrix}^T$. With this formalism the BCS Matsubara Green's function in the SC state is the $2\times2$ matrix:
\be
G_0(\mathbf{k},i\nu_n)\equiv -\int_0^{1/T}\,d\tau\left\langle \mathcal{T}\Psi_\bk(\tau) \Psi^\dagger_\bk(0) \right\rangle e^{i\nu_n\tau}=
\ee
\begin{equation*}
	=\frac{i\nu_n\s_0+\xi_\bk\sigma_3-\Delta\s_1}{(i\nu_n)^2-E_\bk^2}\quad,
\end{equation*}  
where $\nu_n=\pi T(2n+1)$ are fermionic Matsubara frequencies, the 
$\s_a$ are the Pauli matrices, $E_\bk=\sqrt{\xi_\bk^2+\Delta^2}$ and $\D$ is the SC gap, determined as solution of the self-consistent mean-field equation:
\be\label{BCS_EQ}
1=UN\Pi\quad,
\ee
with $N$ the density of states evaluated at the Fermi level and $\Pi=\int_0^{\omega_D}\,d\xi\tanh[E(\xi)/2T]/E(\xi)$. Here $\omega_D$ represents the Debye frequency for the standard phonon-mediated superconductivity, but more generally provides an upper cut-off for the SC interaction for  any pairing mechanism, as e.g. the one provided by the exchange of spin fluctuations in FeSC. 

To introduce the dynamics in the model \eqref{model} we use the path integral formulation, by defining the imaginary-time action for the fermions:
\be\label{FERMI_ACTION}
	S[\Psi,\Psi^\dagger]=\int_0^{1/T}\,d\tau\left[\Psi^\dagger_\bk(\tau)\partial_\tau\Psi_\bk(\tau)+H\right]\quad,
\ee
from which the partition function is given as the functional integral: $\mathcal{Z}=\int\mathcal{D}[\Psi,\Psi^\dagger]e^{-S[\Psi,\Psi^\dagger]}$.

In order to perform the integration over the fermions we use the standard Hubbard-Stratonovich (HS) technique,\cite{nagaosa,depalo_prb99,randeria_prb00,benfatto_prb04} which requires the introduction of bosonic fields to decouple the fermionic interaction terms. In this case, the presence of the two-particle interaction terms $H_P$ and $H_C$ requires to introduce a complex field $h$, which couples to the pairing operator $\Phi_\Delta$ and represents the fluctuations of the SC order parameter around $\Delta$, and a real field, $\rho$, which couples to the density operator $\Phi_\rho$ and represents the density fluctuations. It is
worth noting that the choice of the HS decoupling of an interacting model as in Eq.\ \pref{model}  is not unique, as it has been often discussed in the literature.\cite{depalo_prb99,randeria_prb00,efetov_prb10} However, since we are interested here in deriving the Raman in the SC state, a natural choice for the pairing
term \pref{modelpair} is a decoupling in the SC sector, in order to describe both the SC ground state and the SC fluctuations above it. In addition, as explained in Sec. II, the divergence of the long-range potential $V(\bq)$ plays a crucial role in determining the screening of the Raman response, so we added explicitly the term \pref{modelc}  where the momentum dependence of the density-density interaction, absent in Eq.\ \pref{modelpair}, is taken into account.
Once performed the HS decoupling the action is Gaussian in the fermionic fields, that can be explicitly integrated out. We are then left with the effective action for the HS fields only:
\be
	\mathcal{Z}=\int\mathcal{D}[h,h^\dagger,\rho]e^{-S_{eff}[h,h^\dagger,\rho]}
\ee
\be
	S_{eff}[h,h^\dagger,\rho]=S_{MF}(\Delta)+S_{FL}[h,h^\dagger,\rho]\quad,
\ee
where
\be
S_{MF}(\Delta)=\frac{\Delta^2}{TU}-\mathrm{Tr}\ln\left(-G_0^{-1}\right)\quad.
\ee
 is the mean-field action, that is stationary for $\D$ satisfying the BCS equation \eqref{BCS_EQ}, and $S_{FL}$ is the flucuating action of the HS fields:
 \begin{widetext}
 \be
 	S_{FL}[h,h^\dagger,\rho]=\sum_q\left[\frac{|h(q)|^2}{U}+\frac{|\rho(q)|^2}{2V(\bq)}\right]+
	\sum_{n\ge1}\frac{\mathrm{Tr}\left(G_0\Sigma\right)^n}{n}\quad,
 \ee
 \end{widetext}
 where $q=(i\omega_n,\mathbf{q})$, $\omega_n=2\pi Tn$ are bosonic frequencies and $\Sigma$ is the self-energy of the HS fields. Below the SC critical temperature $T_C\simeq1.13\omega_De^{-1/NU}$ one is always allowed to represent the $h$ field in polar coordinates: $h=|h|e^{i\theta}$. 
 Since we are ultimately interested in the Raman response at $\bf q\ra 0$, we can neglect from the beginning the fluctuations of the amplitude of $h$, since they do not
couple to the phase/density ones in the dynamic limit due to the particle-hole symmetry.\cite{cea_cdw_prb14,cea_prl15} Thus the HS self-energy $\Sigma$ reads:
 \begin{widetext}
 \be
 \lb{self}
 	\Sigma(k,k')=-i\sqrt{T}\rho(k-k')\s_3-\frac{i}{2}\sqrt{T}\theta(k-k')\left[(k-k')_0\s_3-\left(\xi_\bk-\xi_{\bk'}\right)\s_0\right]-
	\ee
	\begin{equation*}
	-
	\frac{T\s_3}{2d}\sum_{q_1,q_2,l}\theta(q_1)\theta(q_2)
	\frac{\partial^2\xi_\bk}{\partial k_l^2}\sin\left(\bq_1/2\right)\sin\left(\bq_2/2\right)\delta(q_1+q_2-k+k')+O\left(\theta^3\right)\quad,
 \end{equation*}
 \end{widetext}
 with $d=2$ the spatial dimension.
 
 By retaining only the harmonic terms we finally obtain the following low-momentum expansion of $S_{FL}$:\cite{depalo_prb99,randeria_prb00,benfatto_prb04}
\bea
S_{FL}&\simeq&\frac{1}{2}\sum_q\left\{
 \left(\frac{1}{V_\bq} - \chi_{\rho\rho}(q) \right) |\rho(q)|^2+\right.\nn\\
&+&\left. \frac{1}{4}\left(-\chi_{\rho\rho}\omega_n^2+D_s\bq^2\right)|\theta(q)|^2 +\right.\nn\\
&-&\left.  \chi_{\rho\rho}(q) \rho(-q) i\omega_n \theta(q) \right\}\quad,
\eea
where $D_s$ is the superfluid stiffness. Since both the density and SC phase carry out a $\sigma_3$ structure in the Nambu space, see Eq.\ \pref{self}, the fermionic susceptibilities that appear as coefficients of the action are all proportional to the charge susceptibility, defined in general as:
\bea
\chi_{\rho\rho}(q)=T\sum_k\mathrm{Tr}\left[G_0(k+q)\s_3G_0(k)\s_3\right], 
\eea
and its $\bq=0$ value is given by Eq.\ \pref{chirhorho}. 

To compute the Raman response function we introduce in the model \eqref{model} a source term $\rho_R$ coupled to the Raman density operator $ \Phi_R(\mathbf{q})\equiv
\sum_{\mathbf{k}\sigma}\gamma(\mathbf{k})c^\dagger_{\mathbf{k}-\mathbf{q}/2,\sigma}c_{\mathbf{k}+\mathbf{q}/2,\sigma}$:
\be
 H\rightarrow H-\sum_\bq\rho_R(-\bq) \Phi_R(\mathbf{q})\quad.
\ee
The dynamic response function can then be obtained as functional derivative with respect to the external field $\rho_R$, see Eq.\ \pref{chirr} above. 
In the effective action formalism the field $\rho_R$ acts as an additional bosonic field in the self-energy $\Sigma$ of Eq.\ \pref{self}, which now becomes:
\be
	\Sigma(k,k')\rightarrow \Sigma(k,k')-\sqrt{T}\rho_R(k-k')\gamma\left[(\bk+\bk')/2\right]\s_3\quad.
\ee
Also the Raman field $\rho_R$ carries a $\sigma_3$ structure in Nambu space, consequence of the fact that the Raman operator is a momentum-modulated density operator. The only difference in the fermionic susceptibilities appearing as coefficients in the effective action is in the $\gamma(\bk)$ factors entering the various bubbles \pref{chizrr}-\pref{chirhorho} of  Eq.\ \pref{sapprox} above, giving the action in the presence of Raman fluctuations. 
Since $V(\bq)\to\infty$ at long-wavelengths, the $\bq=0$ component of \eqref{sapprox} reads:
\bea\label{SAPPROX2}
S_{FL}&=&\frac{1}{2}\sum_q\left\{ |\rho_R|^2\chi^0_{RR}+2i\rho_R\chi_{R\rho}\left[\rho+i\omega_n \theta/{2}\right]-\right.\nn\\
&-&\left.\chi_{\rho\rho}|\rho+i\omega_n{\theta}/{2}|^2\right\},
\eea
where we highlighted that the density and phase fluctuations act as a single field, appearing always as the combination $\rho+i\omega_n{\theta}/{2}$. This is an obvious consequence of the gauge invariance, which allows to reduce the number of degrees of freedom by removing the field $\theta$ via the gauge transformation $\rho+i\omega\theta/2\ra \rho$. In this situation one immediately sees that after integrating out the density fluctuations $\rho$ we are left with the effective action of the source field $\rho_R$ only:
\be
S[\rho_R]=\frac{1}{2}\sum_q |\rho_R|^2\left(\chi^0_{RR}-\frac{\chi_{R\rho}^2}{\chi_{\rho\rho}}\right)\quad,
\ee
from which the functional derivative with respect to $\rho_R$, see Eq.\ \pref{chirr}, leads to the gauge-invariant result \pref{chigi}.  

To prove that the result \pref{chirr} in independent on the presence of long-range Coulomb interaction let us consider again the expression \pref{sapprox} for the effective action when $V(\bq)\ra V$ is replaced by a short-range repulsive potential. In this case the gauge transformation $\rho+i\omega\theta/2\ra \rho$ does not remove the coupling to the phase field. Indeed, after integration of the density field only one recovers the action:
\bea
\lb{srhoonly}
S_{FL}&=&\frac{1}{2}\sum_q\left\{ |\rho_R|^2\left(\chi^0_{RR}-\frac{\chi_{R\rho}^2}{\chi_{\rho\rho}-1/V}\right)-\right.\nn\\
&-&\left.\frac{\chi_{R\rho} \omega_n\theta}{1-V\chi_{\rho\rho}}-\frac{\chi_{\rho\rho}\omega_n^2|\theta|^2/4}{1-V\chi_{\rho\rho}}\right\}.
\eea
In this case, the coefficient of the $|\rho_R|^2$ term coincides with the expression \pref{chisr}, that is manifestely {\em not} gauge invariant. On the other hand, in Eq.\ \pref{srhoonly} the Raman density is still coupled to the phase field. If one then integrates $\theta$ out it is easy to see that the gauge-invariant result \pref{chigi} is once more recovered.

\vspace{1cm}

\section{Derivation of the Raman response for the two-band case}
As a microscopic model for a two-bands superconductor we consider the straight generalization of \eqref{model}:
\bea\label{MODEL_2_BANDS}
H&=&\sum_{\bk,\s,i} \xi_\bk^i c^{i,\dagger}_{\bk,\s}c^i_{\bk,\s} +H_P+H_C\\
H_P&=&-\sum_{i,j,\bq}U_{ij}\Phi^{i,\dagger}_\D(\bq)\Phi^j_\Delta(\bq)\\
H_C&=&\sum_{\bq}V(\bq)\Phi^{\dagger}_\rho(\bq)\Phi_\rho(\bq)
\eea
where $i,j=1,2$ are the band indexes, $\Phi^i_\Delta(\bq)= \sum_\bk c^i_{-\bk+\bq/2,\down}c^{i}_{\bk+\bq/2,\up}$ and  $\Phi_\rho(\bq)= \sum_{\bk,i,\s} c^{i,\dagger}_{\bk-\bq/2,\s}c^i_{\bk+\bq/2,\s}$ are the pairing and density operators, respectively and $\hat U\equiv U_{ij}$ is the matrix of the SC couplings.

At mean-field level, the values of the gaps in each band are given by two coupled self-consistent equations:
\be
\Delta_i=\sum_jU_{ij}\Delta_jN_j\Pi_j\quad,
\ee
with $N_j$ the density of the states of the $j-\mathrm{th}$ band evaluated at the Fermi level.

The Hubbard-Stratonovich technique we used in the single-band model for deriving the effective action of the collective modes can be straightforwardly  generalized to the case of a two-band system, with the foresight of introducing two complex HS fields, $h_1$ and $h_2$, representing the fluctuations of the SC order parameters in each band.

Defining $\theta_i$ the phase of the field $h_i$, the effective action of the phase and density fluctuations reads:
\be
S_{eff}[\theta_1,\theta_2,\rho]=S_{MF}(\Delta_1,\Delta_2)+S_{FL}[\theta_1,\theta_2],
\ee
where:
\be
S_{MF}=\sum_{ij}U^{-1}_{ij}\Delta_i\Delta_j-\sum_i\mathrm{Tr}\ln\left[-G_{0,i}^{-1}\right]
\ee
and
\begin{widetext}
\bea
S_{FL}&\simeq &\frac{1}{2}\sum_q 
\left(\chi^0_{R_1R_1}+\chi^0_{R_2R_2}\right)\left|\rho_R(q)\right|^2+2i\rho_R(-q)\begin{pmatrix}\chi_{R_1\rho_1},\chi_{R_2\rho_2},\chi_{R_1\rho_1}+\chi_{R_2\rho_2}  \end{pmatrix}\begin{pmatrix}\frac{i\omega_n}{2}\theta_1(q)\\\frac{i\omega_n}{2}\theta_2(q)\\\rho(q)\end{pmatrix}+\nn\\
\lb{sfl2}
 &+&
 \lb{sfl2}
\begin{pmatrix}-\frac{i\omega_n}{2}\theta_1(-q),-\frac{i\omega_n}{2}\theta_2(-q),\rho(-q)\end{pmatrix}M(q)\begin{pmatrix}\frac{i\omega_n}{2}\theta_1(q)\\\frac{i\omega_n}{2}\theta_2(q)\\\rho(q)\end{pmatrix},
\eea
\end{widetext}
$M$ being the $3\times3$ matrix:
\begin{widetext}
\bea
\lb{mq}
M(q)=
\begin{pmatrix}
-\chi_{\rho_1\rho_1}+\frac{\kappa+D_{s1}\bq^2}{\omega_n^2}&-\frac{\kappa}{\omega_n^2}&-\chi_{\rho_1\rho_1}\\
-\frac{\kappa}{\omega_n^2}&-\chi_{\rho_2\rho_2}+\frac{\kappa+D_{s2}\bq^2}{\omega_n^2}&-\chi_{\rho_2\rho_2}\\
-\chi_{\rho_1\rho_1}&-\chi_{\rho_2\rho_2}&\frac{1}{V(\bq)}-\chi_{\rho_1\rho_1}-\chi_{\rho_2\rho_2}
\end{pmatrix},
\eea
\end{widetext}

with $\kappa\equiv 8\Delta_1\Delta_2U_{12}/\mathrm{det}U$. Here we defined the
fermionic susceptibilities as a multiband analogous of Eq.s\ \pref{chizrr}-\pref{chirhorho}, so that
\bea
\lb{chizrri}
\chi^0_{R_iR_i}(i\omega_n)&=&-\sum_\bk (\gamma^i_\bk)^2 F^i_\bk(i\omega_n)\\
\lb{chirrhoi}
\chi_{R_i\rho_i}(i\omega_n)&=&- \sum_\bk \gamma^i_\bk F^i_\bk(i\omega_n)\\
\lb{chirhorhoi}
\chi_{\rho_i\rho_i}(i\omega_n) &=& -\sum_\bk  F^i_\bk(i\omega_n)
\eea
where
\be
\lb{fki}
F^i_\bk(i\omega_n)= 4\Delta_i^2
\frac{\tanh(E_{i,\mathbf{k}}/2T)}{E_{i,\mathbf{k}}\left[4E_{i,\mathbf{k}}^2-(i\omega_n)^2\right]},
\ee

As it has been discussed in Ref.\ \cite{marciani_prb13}, in the case of dominant interband pairing the derivation of Eq.\ \pref{sfl2} is more involved, since in this case the matrix of the SC couplings $\hat U$ admits a negative eigenvalue corresponding to the presence of an antibonding SC channel. In this case one can still implement the Hubbard-Stratonovich decoupling by introducing first a combination of the fermionic fields that allows one to impose a vanishing saddle-point value of the antibonding channel. Afterwards one can express back the fluctuations in terms of the collective modes in each band, obtaining the structure \pref{sfl2} of the collective-mode action. 

In the limit $\bq=0$ one can notice that $M(i\omega_n)$ is always singular, having $(1,1,-1)$ as eigenvector corresponding to the zero eigenvalue, as one can immediately check by summing the lines of the matrix \pref{mq}. This means that the description in terms of three degrees of freedom $\theta_1,\theta_2,\rho$ is redundant and we can invoke the gauge invariance to remove one of them. To show this formally, it turns useful to introduce the new variables:
\bea
\theta_G&=&\frac{\theta_1+\theta_2}{2}\\
\theta_L&=&\frac{\theta_1-\theta_2}{2}\\
\tilde{\rho}&=&\rho+\frac{i\omega_n}{2}\theta_G
\eea
where the subscripts $G$ and $L$ denote the Goldstone and Leggett phase mode, respectively, while $\tilde{\rho}$ defines a gauge transformation of the field $\rho$.

One can easily check that in the new frame the matrix $M$ becomes:
\begin{widetext}
\be
M(q)=
\begin{pmatrix}
 0&0&0\\
0&-\chi_{\rho_1\rho_1}-\chi_{\rho_2\rho_2}+4\frac{\kappa}{\omega_n^2}&-\chi_{\rho_1\rho_1}+\chi_{\rho_2\rho_2}\\
0&-\chi_{\rho_1\rho_1}+\chi_{\rho_2\rho_2}&-\chi_{\rho_1\rho_1}-\chi_{\rho_2\rho_2}
\end{pmatrix}.
\ee
\end{widetext}
Then the field associated to the Goldstone mode $\theta_G$ does not couple to any other field and $S_{FL}$ reduces to a functional of the fields $\theta_L$ and $\tilde{\rho}$ only:
\begin{widetext}
\be
\label{S_FL_2_BANDS}
	S_{FL}\simeq\frac{1}{2}\sum_n\left\{
	\left(\chi_{R_1R_1}+\chi_{R_2R_2}\right)\left|\rho_R(i\omega_n)\right|^2+2i\rho_R(-i\omega_n)\begin{pmatrix}\chi_{R_1\rho_1}-\chi_{R_2\rho_2},\chi_{R_1\rho_1}+\chi_{R_2\rho_2}  \end{pmatrix}\begin{pmatrix}\frac{i\omega_n}{2}\theta_L(i\omega_n)\\\tilde{\rho}(i\omega_n)\end{pmatrix}+\right.
\ee
\begin{equation*}\left.
	\begin{pmatrix}-\frac{i\omega_n}{2}\theta_L(-i\omega_n),\tilde{\rho}(-i\omega_n)\end{pmatrix}
	\tilde{M}(i\omega_n)	\begin{pmatrix}\frac{i\omega_n}{2}\theta_L(i\omega_n)\\\tilde{\rho}(i\omega_n)\end{pmatrix}\right\}
\end{equation*}
\end{widetext}
where $\tilde{M}$ is the $2\times2$ matrix:
\be
\tilde{M}(i\omega_n)=\begin{pmatrix} -\chi_{\rho_1\rho_1}-\chi_{\rho_2\rho_2}+4\frac{\kappa}{\omega_n^2}&-\chi_{\rho_1\rho_1}+\chi_{\rho_2\rho_2}\\
-\chi_{\rho_1\rho_1}+\chi_{\rho_2\rho_2}&-\chi_{\rho_1\rho_1}-\chi_{\rho_2\rho_2}\end{pmatrix}\quad,
\ee
which becomes singular at $i\omega_n=\omega_L$, where $\omega_L$ is the Leggett frequency, given by the solution of:
\bea
\omega_L^2&=&F_L(\omega_L)\nn\\
F_L(i\omega_n)&\equiv& -\kappa\frac{\chi_{\rho_1\rho_1}(i\omega_n)\chi_{\rho_2\rho_2}(i\omega_n)}{\chi_{\rho_1\rho_1}(i\omega_n)+\chi_{\rho_2\rho_2}(i\omega_n)}=\nn\\
&\simeq&\kappa\frac{N_1F_1(i\omega_n)+N_2F_2(i\omega_n)}{N_1N_2F_1(i\omega_n)F_2(i\omega_n)},
\eea
that coincides with Eq.\ \pref{eqleg} above. 

From Eq.\ \pref{S_FL_2_BANDS} one immediately sees that the coupling to the Legget $\theta_L$ and to the charge $\rho$ fluctuations is dictated by the same susceptibilities $\chi_{R_i\rho_i}$, even though combined with different signs. If one integrates out {\em only} the density modes it is straightforward to see that the coefficient of the $|\rho_R|^2$ field becomes Eq.\ \pref{chingi}, as stated in Ref.\ [\onlinecite{boyd_prb09,mazin_prb10,blumberg_prb10}]. Thus, for parabolic bands having equal DOS and opposite character $\chi_{R_1\rho_1}=-\chi_{R_2\rho_2}$ and the coupling to the density mode cancels out. However, the coupling to the Leggett mode cannot be removed, since it is maximum under the same condition. This is expected on physical ground, since in this case the Raman operator is proportional to relative density fluctuations between the two bands, that are conjugated to the Leggett fluctuations.  

To obtain the full Raman response function one should then integrate {\em both} the fields $\theta_L$ and $\tilde{\rho}$, obtaining the effective action of $\rho_R$ only:
\be
	S[\rho_R]=\frac{1}{2}\sum_q |\rho_R|^2\chi_{RR}\quad,
\ee
with:
\begin{widetext}
\be\label{CHI_TRUE_APP}
	\chi_{RR}=\left(\chi_{R_1R_1}+\chi_{R_2R_2}\right)+\begin{pmatrix}\chi_{R_1\rho_1}-\chi_{R_2\rho_2},\chi_{R_1\rho_1}+\chi_{R_2\rho_2}  \end{pmatrix}\tilde{M}^{-1}\begin{pmatrix}\chi_{R_1\rho_1}-\chi_{R_2\rho_2}\\\chi_{R_1\rho_1}+\chi_{R_2\rho_2}  \end{pmatrix}=
\ee
\begin{equation*}
	=\left(  \chi_{R_1R_1}+  \chi_{R_2R_2}  \right)-\frac{ (i\omega_n)^2\left(	\chi_{\rho_1\rho_1}\chi_{\rho_2R_2}^2+\chi_{\rho_2\rho_2}\chi_{\rho_1R_1}^2\right)+\kappa\left(\chi_{\rho_1R_1}+\chi_{\rho_2R_2}\right)^2  }{(i\omega_n)^2\chi_{\rho_1\rho_1}\chi_{\rho_2\rho_2}+\kappa\left(\chi_{\rho_1\rho_1}+\chi_{\rho_2\rho_2}\right)}\quad,
\end{equation*}
\end{widetext}
that gives back Eq.\ \pref{chitrue}.

\end{appendix}



\begin{thebibliography}{99}
\bibitem{note_raman}
Here we refer to the 1Fe BZ notation, that has been adopted in Ref.\ [\onlinecite{gallais_prl16,hackl_prl13,hackl_prx14}]. Notice that instead in Ref.\ [\onlinecite{blumberg_prb16}] the notation is for the 2Fe BZ, that is rotated by $\pi/4$, so that the relevant Raman channel becomes $B_{2g}$.
\end{thebibliography}
\end{document}